\documentclass[a4paper,american,prl,twocolumn]{revtex4-1}
\usepackage[T1]{fontenc}
\usepackage[latin9]{inputenc}
\usepackage{amsmath}
\usepackage{graphicx}

\makeatletter

\pdfpageheight\paperheight
\pdfpagewidth\paperwidth

\usepackage[colorlinks = true,citecolor=blue,linkcolor=blue,urlcolor=blue,pdfborder={0 0 0}]{hyperref}

\makeatother

\usepackage{babel}
\begin{document}
\title{Dynamical Network Models of the Turbulent Cascade}
\author{Ö. D. Gürcan}
\affiliation{Laboratoire de Physique des Plasmas, CNRS, Ecole Polytechnique, Sorbonne
Université, Université Paris-Saclay, Observatoire de Paris, F-91120
Palaiseau, France}
\begin{abstract}
Cascade models based on dynamical complex networks are proposed as
models of turbulent energy cascade. Taking a simple shell model as
the initial regular lattice with only nearest neighbor interactions,
small world network models are constructed by adding or replacing
some of the existing local interactions by nonlocal ones. The models
are then evolved over time, both by solving for the shell variable
for velocity using an arbitrary network generalization of the shell
model evolution and by rewiring the network each time from the original
lattice in regular time intervals. This results in a more intermittent
time evolution with larger variations of the wave-number spectrum.
It also results in an actual increase in intermittency as computed
from the fitted exponents of structure functions as computed from
these models. It appears that the intermittency increases as the ratio
of random nonlocal connections to local nearest neighbor connections
increases.
\end{abstract}
\maketitle

\section{Introduction}

Turbulence is a duality of chaotic disorder and hierarchical organization
across a large range of scales in the evolution of a fluid. This aspect
of turbulence is shared with many other self-organizing complex systems,
that are commonly described using networks, such as the internet\citep{albert:99},
the brain\citep{bullmore:09}, the public transport infrastructure\citep{seaton:04},
and the economy\citep{kirman:97} - to give a few examples.

The three dimensional incompressible fluid, described by the Navier-Stokes
equation, mixed at large scales by an external forcing and dissipated
at small scales due to molecular viscosity, provide the canonical
example of the turbulent cascade. The idealized process of self-similar
energy transfer from large scales where the energy is injected to
small scales, where it is dissipated can be described using various
simplified models including shell models\citep{biferale:03}, differential
approximations\citep{leith:67} or closures\citep{lesieur:book}.
While the study of turbulence has a long history\citep{frisch}, and
networks are ubiquitous in modern nonlinear science\citep{barabasi:11},
the connection between the two is an emerging field with many open
questions\citep{taira:16,gurcan:20}.

Network theory is as much about spreading or flow of various constituents
such as information, ideas or pathogens within a given (or evolving)
network structure, as it is about the topology of the network itself.
Flow of packages through the internet, people through the public transport
system\citep{lee:08}, or the spreading of financial crises through
the global economy\citep{kali:10} or a deadly virus across a network
of human contacts\citep{kelling:05} are all examples of such phenomena
that fall under the umbrella term of percolation in complex networks\citep{dorogovtsev:08,watts:02}.
The turbulent cascade of energy in a complex network representing
the wavenumber domain, fits right in with the rest of these examples.
However there is a key difference in the turbulent problem: the interactions
are between three nodes instead of two, since they are produced by
triadic interactions. In this sense the turbulent cascade takes place
one a network with \emph{``three body''} interactions\citep{neuhauser:20}.

In this spirit, we propose a concrete working example in the form
of a model of the turbulent cascade using dynamical complex networks
by generalizing the GOY model\citep{ohkitani:89} to a percolation
model on a complex, small world network with long range interactions.
This allows the exploration of different strategies of random rewiring
of a regular lattice in order to form nontrivial small-world networks\citep{newman:03}
on which the turbulence is then allowed to develop. Two different
rewiring strategies based on Watts-Strogatz and Newman-Watts are discussed
and dynamical network models are considered where the network is regularly
rewired. Energy cascade is described on top of this evolving small
world network using a simple shell model-like evolution of the observables
$u_{n}\left(t\right)\equiv\sqrt{2\int_{k_{n}}^{k_{n+1}}E\left(k\right)dk}$.
While this would probably appear quite unphysical to a specialist
in turbulence, since the turbulence can be described nicely on a constant
regular grid with deterministic equations, its power comes from the
additional degree of freedom -i.e. that of the network topology- that
it provides, which allows us to represent part of what has been lost
in the reduction leading up to the shell model at a very modest cost
-the solution of the network model is not slower than the GOY model-. 

Shell models are closely linked to the concept of spectral reduction\citep{bowman:99},
which basically amounts to reducing the regular spectral domain to
a smaller set of regions. When the full spectral domain is thus reduced,
detailed phase relations between regions are lost. The shell model
does have a complex phase that evolves, but this has almost nothing
to do with the actual phase of the full system. The direction of energy
transfer at a given instant, or its efficiency depends on these phase
relations. If the phases are aligned between two regions, the energy
can be transferred efficiently, while if they are out of phase, there
may be no energy transfer. While the evolution of phases is deterministic
in the full system, its chaotic and usually irregular. Therefore its
effect on a shell model like reduction can be represented plausibly
by connections being turned on and off randomly. Note that on top
of the connection being turned on the phase relations of the shell
model itself still has to be satisfied for the energy transfer to
take place. For a real physical problem, phase relations may be random
or regular, for instance as in the case of weak wave turbulence\citep{newell:11}.
In such a case, the network topology may be constructed respecting
the dispersion relation of the underlying waves, and the network may
be used to represent those 'enhanced' connections between disparate
scales due to resonant interactions. If the statistics of those phenomena
are well represented by the evolution of the network, its time evolution
may correspond better to the time evolution of the unreduced system.

The rest of the paper is organized as follows. In section \ref{sec:swnm}
we introduce the basic small world network paradigm for turbulence
using a generalization of the GOY model. In sections \ref{subsec:ws}
and \ref{subsec:nw} we lay out the strategies for constructing Watts-Strogatz
and Newman-Watts models respectively, while in \ref{subsec:bnm} we
discuss the bipartite network perspective using two sets of nodes
corresponding to wavenumbers and triads. In section \ref{sec:nr}
we provide the numerical results of all these different models compared
to the basic GOY model. Section \ref{sec:conc} is conclusion.

\section{Small World Network Shell Models\label{sec:swnm}}

Consider a shell model of turbulence \citep{biferale:03} with arbitrary
range interactions for three dimensional turbulence. Using a set of
wave-vectors $k_{n}=k_{0}g^{n}$, where $g$ is the logarithmic scaling
factor (usually $g=2$), the model can be written as follows:
\begin{align}
\partial_{t}u_{n}=i\alpha_{m}\bigg[ & a_{n}^{m}u_{n+m}^{*}u_{n+m+1}^{*}+b_{n}^{m}u_{n+1}^{*}u_{n-m}^{*}\nonumber \\
 & +c_{n}^{m}u_{n-1}^{*}u_{n-1-m}^{*}\bigg]\label{eq:goy}
\end{align}
where the interaction coefficients can be written as $a_{n}^{m}=M_{n,n+m,n+m+1}$,
$b_{n}^{m}=M_{n,n-m,n+1}$ and $c_{n}^{m}=M_{n,n-1-m,n-1}$ with:

\begin{equation}
M_{n,\ell,\ell'}=\begin{cases}
k_{\ell}+k_{\ell'} & \text{if }n<\ell\\
-\left(\left(-1\right)^{n-\ell}k_{\ell}+k_{\ell'}\right) & \text{if }\text{\ensuremath{\ell<n}}
\end{cases}\label{eq:Mnlm}
\end{equation}
Here $m$ is the range of interaction (i.e. $m=1$ gives us the usual
GOY model), and $\alpha_{m}$ is the average contribution from the
geometric factor, which we take as $\alpha_{m}=g^{-m}$ (see for example
\citep{plunian:07}). Note that $\ell<\ell'$ is assumed in (\ref{eq:Mnlm}).
This allows conservation of energy 
\[
E=\sum_{n}u_{n}^{2}
\]
 and helicity 
\[
H=\sum_{n}\left(-1\right)^{n}k_{n}^{-1}u_{n}\;\text{,}
\]
since 
\[
M_{n,n+m,n+m+1}+M_{n+m,n,n+m+1}+M_{n+m+1,n,n+m}=0
\]
 and 
\begin{align*}
k_{n}M_{n,n+m,n+m+1}+\left(-1\right)^{m}k_{n+m}M_{n+m,n,n+m+1}\\
+\left(-1\right)^{m+1}k_{n+m+1}M_{n+m+1,n,n+m} & =0\;\text{.}
\end{align*}
\begin{figure}
\begin{centering}
\includegraphics[width=0.7\columnwidth]{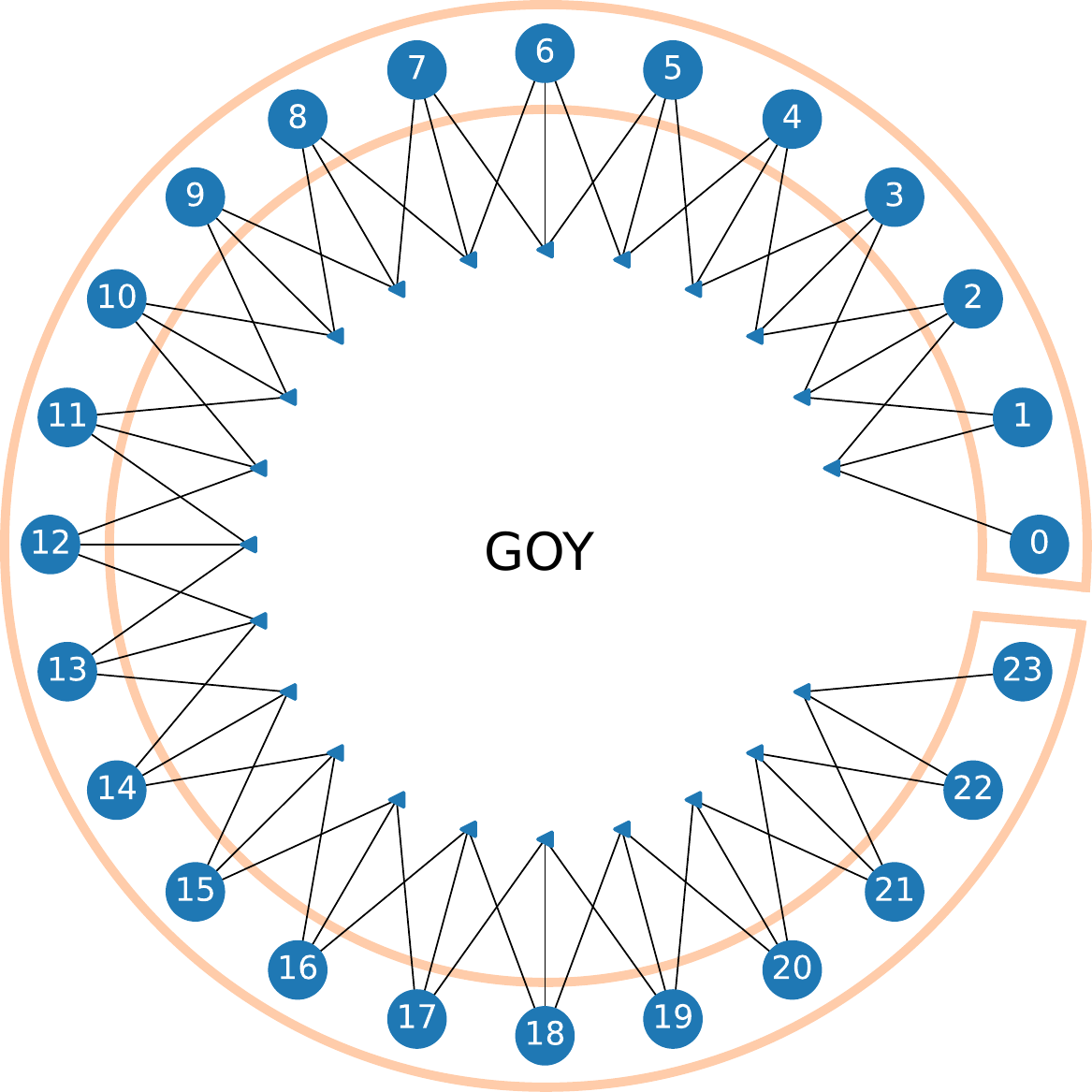}
\par\end{centering}
\caption{\label{fig:goy-1}The regular lattice of the GOY model. Apart from
the end nodes, all the nodes are connected to three triads and each
triad is connected to the three closest nodes. The lattice is shown
in a circular representation in order to save space and connect to
earlier works.}
\end{figure}
The three terms in (\ref{eq:goy}) come from the three triadic interactions
that have the same form but shifted with respect to one another. The
first term proportional to $a_{n}^{m}$ describes the interaction
between the shells $n$, $n+m$ and $n+m+1$. The last two terms proportional
to $b_{n}^{m}$ and $c_{n}^{m}$ are the same interaction but shifted
by $-m$ and $-m-1$ respectively. This allows us to consider different
terms in the equation in terms of undirected triadic interactions.
For example a single triadic connection $\left(n,n+m,n+m+1\right)=\left(1,4,5\right)$
introduces one term in the equation for $u_{1}$ with the coefficient
$a_{1}^{3}=M_{145}$, one term in the equation for $u_{4}$ with the
coefficient $b_{4}^{3}=M_{415}$ and finally one term in the equation
for $u_{5}$ with the coefficient $c_{5}^{3}=M_{514}$.

The standard GOY model corresponds to $m=1$, which represents \emph{``nearest
neighbor''} interactions shown in figure \ref{fig:goy-1}. If we
choose $g=2$, and absorb a prefactor $6$ into the arbitrary constant
$\alpha$, (\ref{eq:Mnlm}) gives $a_{n}=k_{n}$, $b_{n}=-k_{n-1}/2$
and $c_{n}=-k_{n-2}/2$ of Ref. \citealp{ohkitani:89}. 
\begin{figure}
\begin{centering}
\includegraphics[width=0.98\columnwidth]{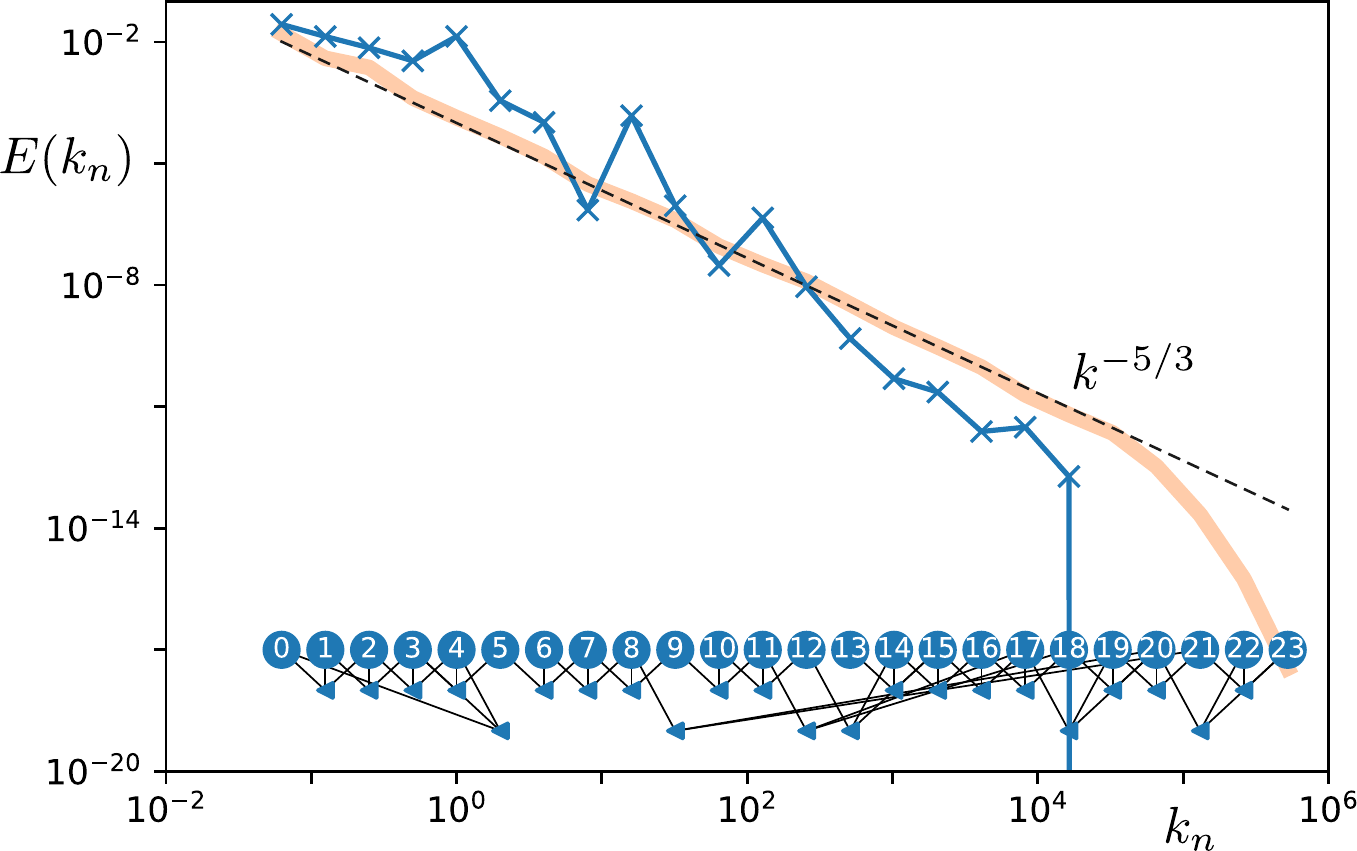}
\par\end{centering}
\caption{\label{fig:ws_spec}Wave number spectrum for the WS network, generated
with $p=0.4$, $p_{f}=0.5$, compared with the GOY model shown with
thick orange (if in color) line. Nodes $5$, $9$, $10$, $12$ and
$13$ are missing connections which results in formation of peaks
or wells. The fact that energy can go to dissipative range through
non-local connections makes the spectrum fall off rapidly at around
$10^{4}$. Here $N=24$, $k_{0}=2^{-4}$, $\nu=10^{-8}$ and $f_{n}=\left(\delta_{n1}\xi_{1}+\delta_{n2}\xi_{2}\right)10^{-2}$,
where $\xi_{i}$ are random variables with a correlation time of $10^{-2}$.
The spectra are integrated up to $t=5\times10^{3}$ and average over
$t=\left[3-5\right]\times10^{3}$ are shown.}
\end{figure}

\subsection{Watts-Strogatz model\label{subsec:ws}}

There is a total of $N-2$ triads in the regular lattice of the GOY
model with $N$ nodes. In order to create a partially randomized network
with non-local interactions, we go over this list of triads and replace
the local triad $\left(n,n+1,n+2\right)$ with a nonlocal one, with
either a forward, {[}i.e. $\left(n,n+m,n+m+1\right)${]} or a backward
{[}i.e. $\left(n,n-m,n-1\right)${]} coupling with a probability $p$,
where $m$ itself is a random number between $2$ and $N-n-2$ for
the forward or between $3$ and $n-2$ for the backward coupling.
We can choose the interaction to be forward with a probability $p_{f}$.
This basic algorithm is very similar to the one described by Watts
and Strogatz in order to build small world networks\citep{watts:98}
(so we call it the Watts-Strogatz model or WS for short), except that
the topology of the initial lattice is not really a ring, and the
connections are not lines, but triadic interactions. 

Having the list of triads thus revised, we can recompute the list
of interactions $\mathbf{i}_{n}=\left\{ \ell',\ell''\right\} $ and
the interaction coefficients (i.e. weights) $M_{n,\ell,\ell'}$ for
each node using the triads that connect to it. The idea is to go over
the list of triads, and for example when treating the triad $\left(n,n+m,n+m+1\right)$,
add the three interactions $\mathbf{i}_{n}=\left\{ n+m,n+m+1\right\} $,
$\mathbf{i}_{n+m}=\left\{ n,n+m+1\right\} $ and $\mathbf{i}_{n+m+1}=\left\{ n,n+m\right\} $
to the list of interactions, with the corresponding interaction coefficients
$M_{n,n+m,n+m+1}$, $M_{n+m,n,n+m+1}$ and $M_{n+m+1,n,n+m}$ respectively.
In the end, one may have nodes with more or less connections than
three (the original number of connections of each node in the GOY
model), and these connections may be local or nonlocal, but since
the contributions from each triad to all its three nodes are always
considered, the conservation laws are automatically respected. The
model goes from the regular shell model with local interactions for
$p=0$ to a cascade model with random scale interactions for $p=1$.
In contrast, $p_{f}$ does not play an important role on network topology,
so one could pick all the connections to be forward without loss of
generality.
\begin{figure}
\begin{centering}
\includegraphics[width=0.7\columnwidth]{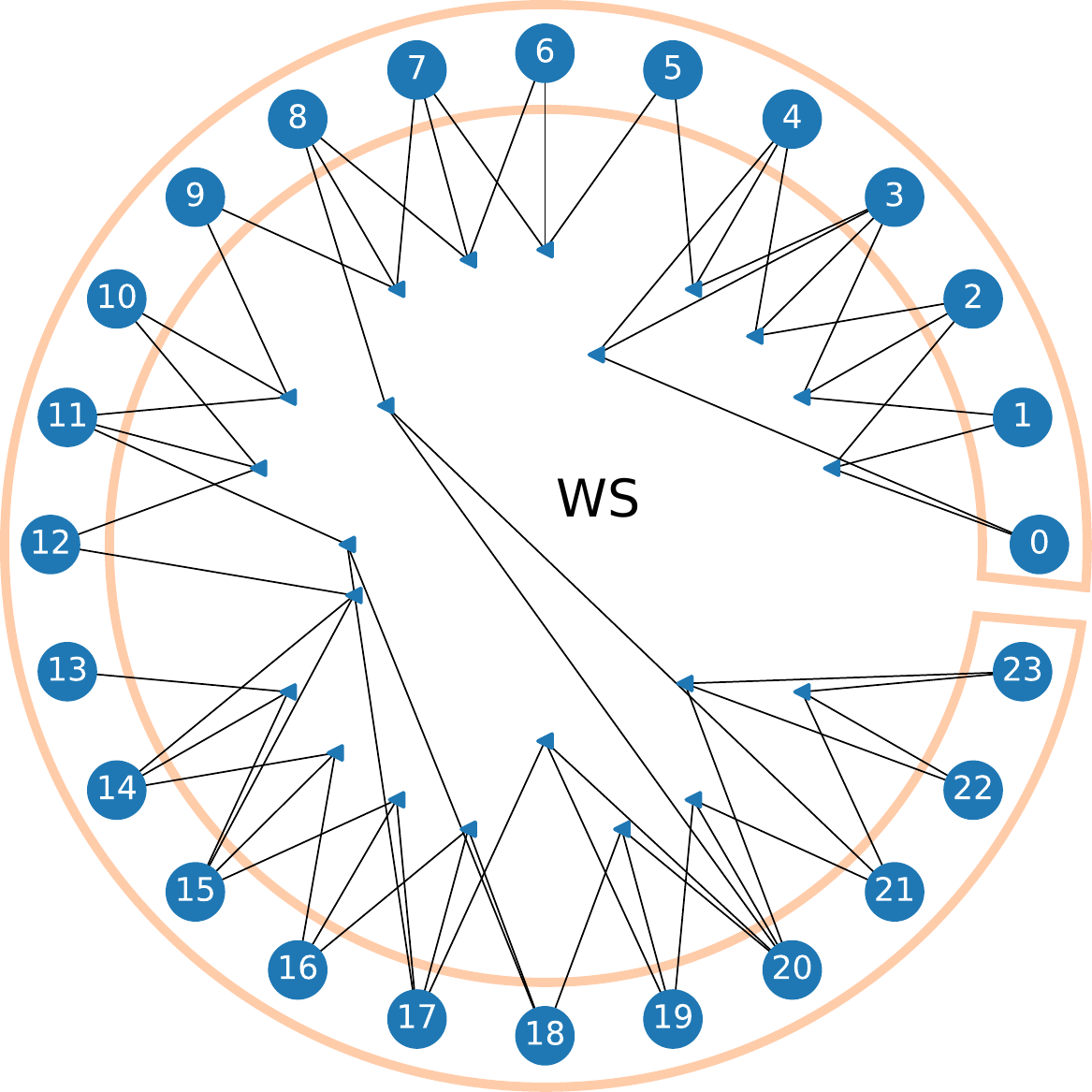}
\par\end{centering}
\caption{\label{fig:ws}The small world network of the WS model, generated
with $p=0.4$,$p_{f}=0.5$ also shown in Figure \ref{fig:ws_spec}
with less than three connections at nodes $5$, $9$, $10$, $12$
and $13$.}
\end{figure}

Once the network is constructed, time evolution of the node variables
$u_{n}$ can be written as:
\begin{align}
\left(\partial_{t}+\nu k_{n}^{2}\right)u_{n} & =\sum_{\ell',\ell''=\mathbf{i}_{n}}M_{n,\ell,\ell'}u_{\ell}^{*}u_{\ell'}^{*}\;+f_{n}\text{,}\label{eq:nweq}
\end{align}
where $\mathbf{i}_{n}$ is the list of interaction pairs for the $n$th
node, and $M_{n,\ell,\ell'}$ are the interaction coefficients, $\nu$
is kinematic viscosity and $f_{n}$ is (localized and random) forcing.

A static network that is integrated for a certain number of time steps
is not a particularly interesting exercise. In particular since the
result relies on initialization and how the network is wired. The
resulting spectrum is a considerably rugged version of the shell model
one, as seen in fig. \ref{fig:ws_spec}, with barriers around nodes
that are missing connections. Furthermore each time the network is
rewired, the details of how it deviates from the regular shell model
would change. Figure \ref{fig:ws} shows the particular wiring in
more detail that leads to the spectrum shown in figure \ref{fig:ws_spec}.
Notice that some nodes are missing connections, and the energy has
difficulty going through those. 

A more realistic approach is to rewire the network in regular time
intervals (i.e. $\Delta t$) as the system evolves. If we run such
a model for a long enough time $t\gg\Delta t$ we can obtain good
statistics. Various interesting problems related to shell models,
such as intermittency etc. can also be studied in this formulation.
Note that in order to not completely randomize the network in a few
time steps, we apply the WS strategy on the original network and not
the modified one at $t$ in order to obtain the network structure
at $t+\Delta t$. The results for this dynamical network formulation
using the WS strategy can be seen in section \ref{sec:nr} along with
results for the other strategies.

\subsection{Newman-Watts model\label{subsec:nw}}

\begin{figure}
\begin{centering}
\includegraphics[width=0.98\columnwidth]{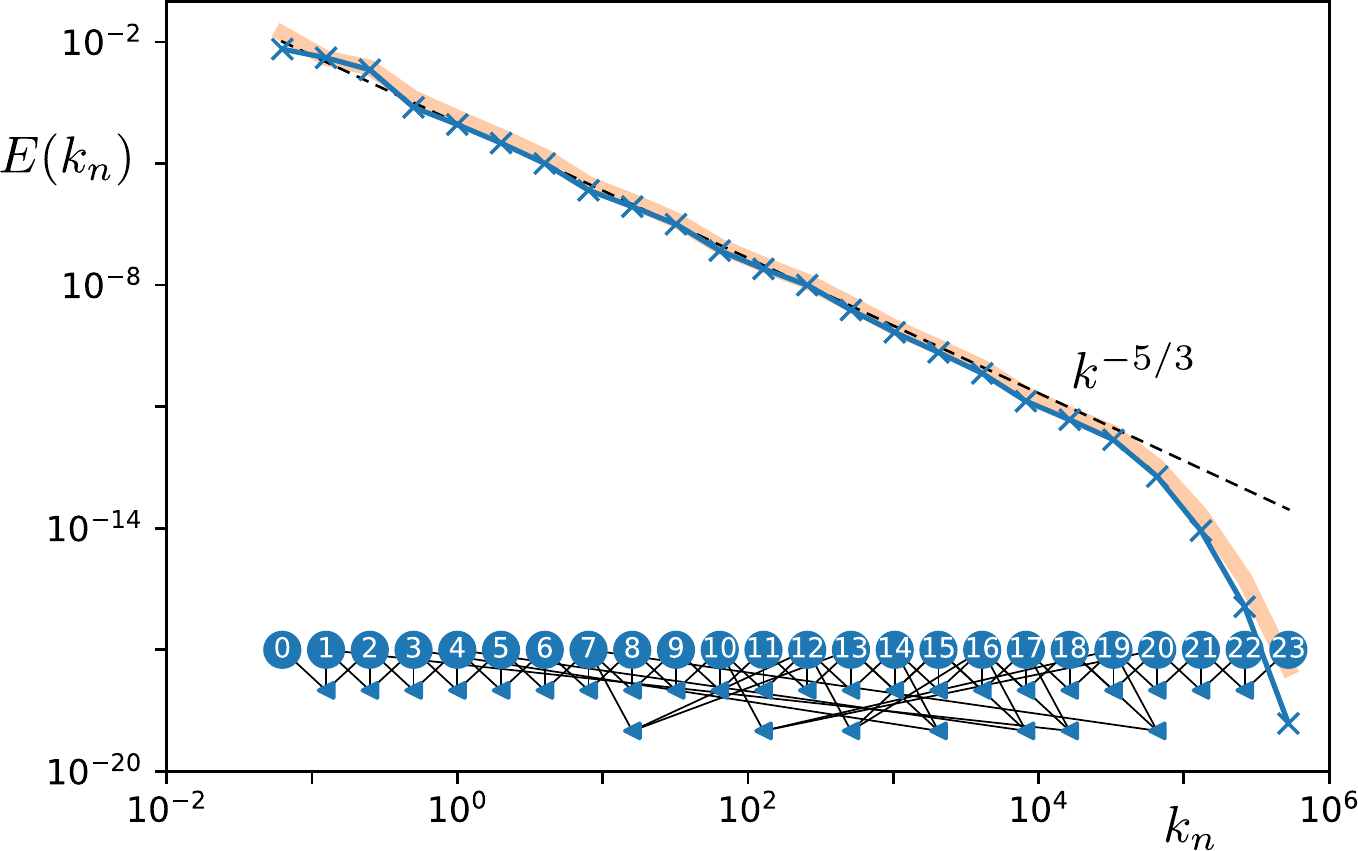}
\par\end{centering}
\caption{\label{fig:nw_spec}Wave number spectrum for the NW network, generated
with $p=0.4$, $p_{f}=1.0$ compared with the GOY model. All the nodes
have at least $3$ connections therefore no barriers appear. Also
the additional connections dissipative range make the spectrum fall
off a bit faster. Parameters and run times are the same as in figure
\ref{fig:ws_spec}.}
\end{figure}

\begin{figure}
\begin{centering}
\includegraphics[width=0.7\columnwidth]{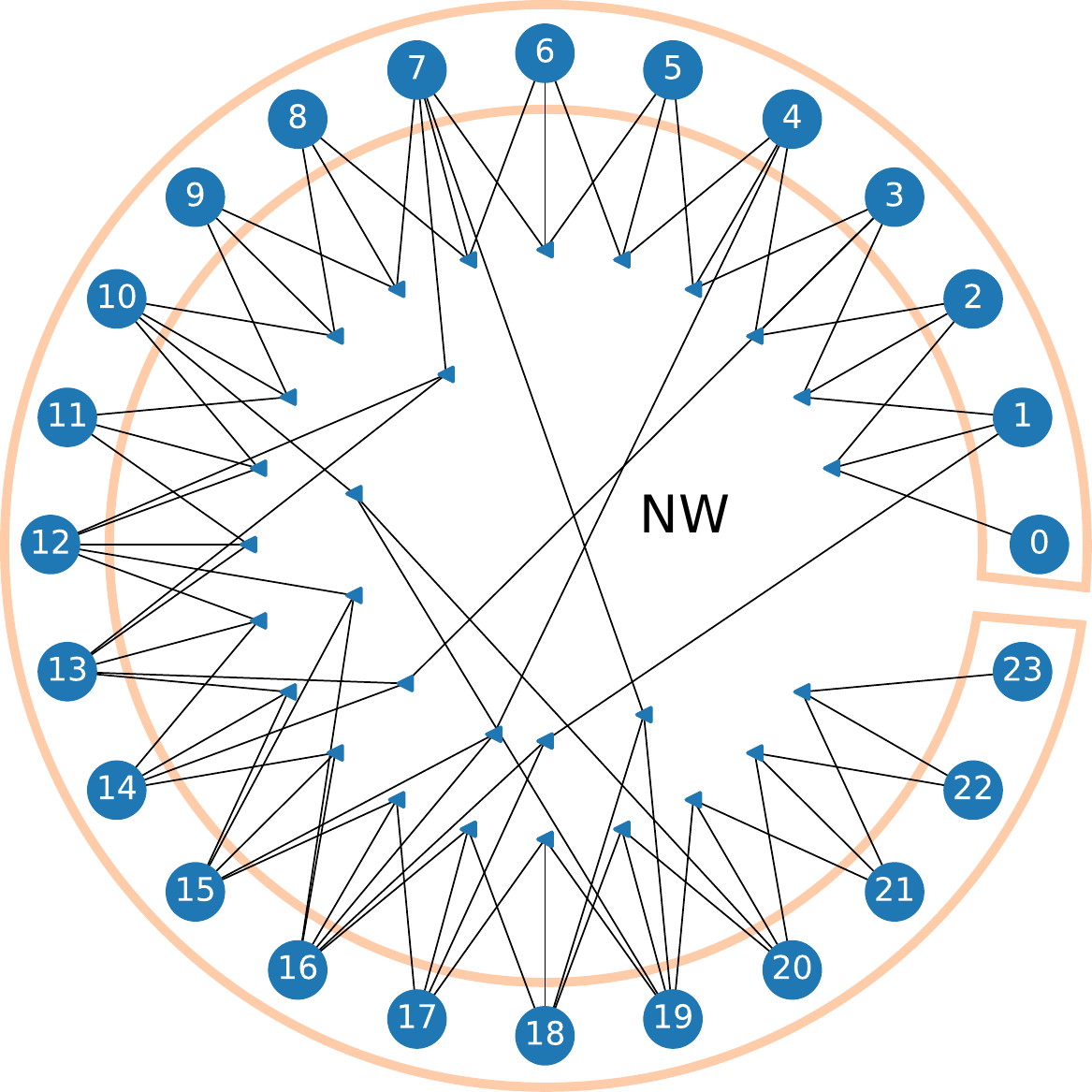}
\par\end{centering}
\caption{\label{fig:nw}The small world network of the NW model, generated
with $p=0.4$, $p_{f}=1.0$, which is the same network as the one
in Figure \ref{fig:nw_spec}.}
\end{figure}

Newman and Watts proposed an alternative algorithm for constructing
a similar partially randomized network from a regular initial lattice\citep{newman:92}.
It translates to shell models as \emph{adding} a non-local triad instead
of replacing the local one as in WS, with, $m$, $p$ and $p_{f}$
having the same roles as before. We call this, the Newman-Watts strategy
or NW for short. The steady-state wave-number spectrum on a network
obtained by this algorithm is shown in figure \ref{fig:nw_spec},
where the network itself is shown in figure \ref{fig:nw}. Note that
since the algorithm simply adds connections and the interaction coefficients
for those nonlocal connections go down as $g^{-m}$, the result is
very similar to GOY. 

The primary advantage of the NW is that it keeps the basic structure
of the underlying regular local lattice. This allows the basic local
transfers to always be present, giving a smoother steady state spectrum.
In any case, the more relevant formulation of the model is not a single
network instance but a dynamically rewired one, whose results are
shown below in section \ref{sec:nr}.

\subsection{Bipartite Networks of Wave-numbers and Triads for Describing Turbulence\label{subsec:bnm}}

The discussion of dynamical complex network models above is based
on interactions between nodes and pairs, where each interaction is
represented by a triad, and we talk about nodes that are connected
to triads. The graphs of networks in figures \ref{fig:goy-1},\ref{fig:ws}
and \ref{fig:nw} show these triads explicitly. This is actually a
hint at the underlying nature of networks that appear in spectral
description of turbulence. These networks with three body interactions
can also be represented as bipartite networks\citep{filho:18} that
exclusively connect two separate kinds of nodes \emph{``wave-number
nodes''} representing wave-number domains and \emph{``triad nodes''}
representing triadic interactions, with the additional constraint
that each triad has three connections. This perspective allows us
to transform networks where nodes are connected to pairs, into the
simpler and well known class of bipartite networks, and to ask common
questions in network topology such as average distance, clustering
coefficients or degree distributions. In particular, using the bipartite
network but focusing on the wave-number nodes, we can construct a
projected simple (or multi or weighted) graph network as discussed
in Ref. \citealp{filho:18}, where the nodes in the projected network
are connected only if they are both connected to the same triad.

\begin{figure}
\begin{centering}
\includegraphics[width=0.8\columnwidth]{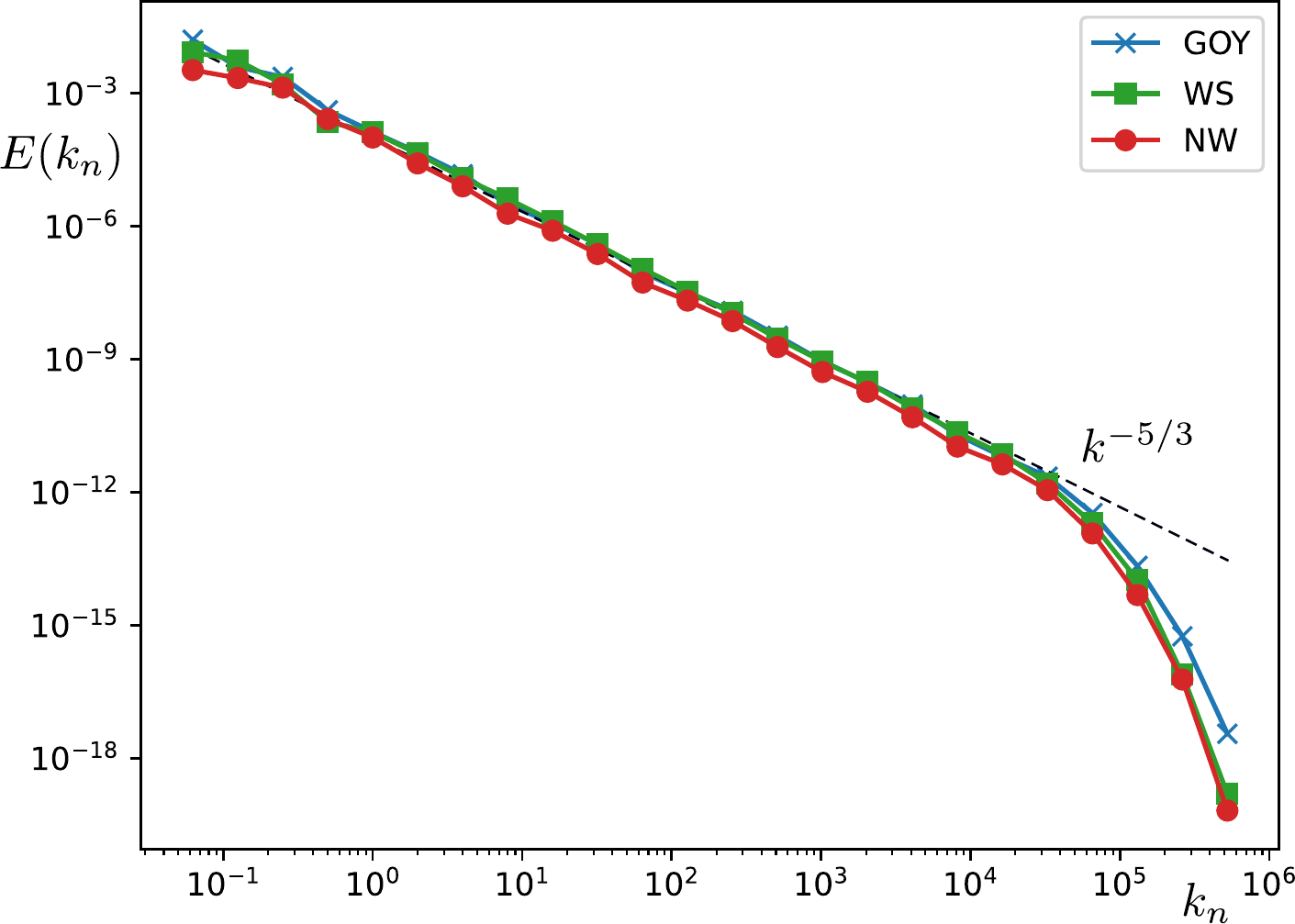}
\par\end{centering}
\caption{\label{fig:specs}The resulting steady state spectra from the dynamical
complex network models WS and NW compared with that of the GOY model,
showing that all three models basically capture the $k^{-5/3}$ spectrum
that we expect, while NW is very slightly lower in amplitude as opposed
to the other two probably as a result of its extra connections, and
therefore higher transfer efficiency. The parameters for these runs
are discussed in the text.}
\end{figure}

\section{Numerical Results\label{sec:nr}}

Here we focus on the results from the dynamical network models discussed
above, which are rewired according to either WS or NW in regular intervals
of $\Delta t$. Unlike the static cases, there is no big difference
between the two in terms of their steady state spectra as shown in figure \ref{fig:specs}, since an evolving
network moves its barriers around, and as a result, allows energy
transfer, more easily. The results shown in this section uses the
parameters $N=24$, $k_{0}=2^{-4}$, $\nu=10^{-8}$ and $f_{n}=\left(\delta_{n1}\xi_{1}+\delta_{n2}\xi_{2}\right)10^{-2}$,
where $\xi_{i}$ are random variables with a correlation time of $10^{-2}$.
The spectra are integrated up to $t=5\times10^{3}$ using an adaptive
fourth order Runge Kutta solver\citep{gurcan_dycon}, and when steady
state results are shown they are usually averaged over $t=\left[3-5\right]\times10^{3}$.

\ref{fig:ws_spec}. 

\begin{figure}
\begin{centering}
\includegraphics[width=0.98\columnwidth]{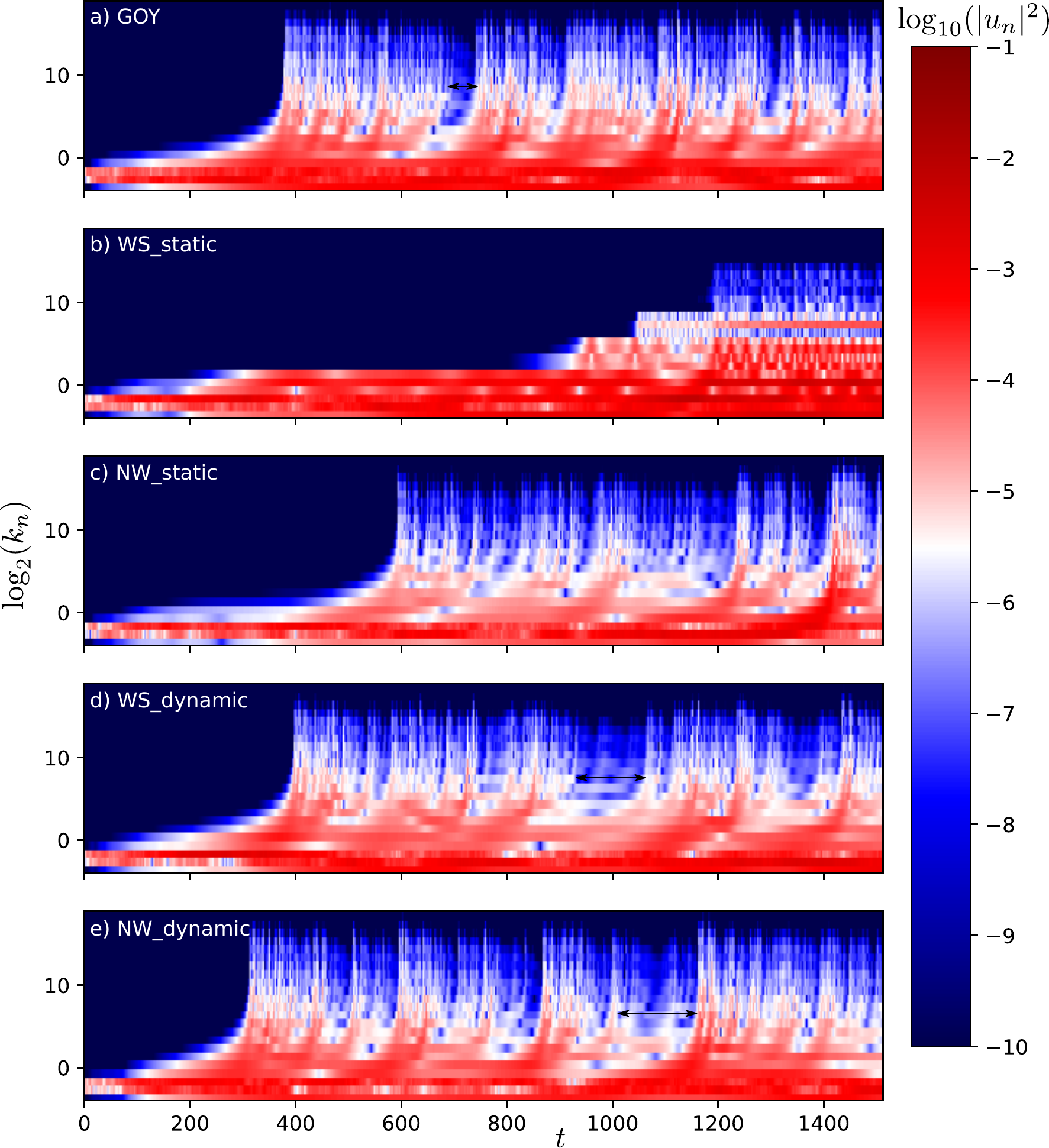}
\par\end{centering}
\caption{\label{fig:evol}Time evolution of the wave number spectrum, up to
$t=1500$ for a) GOY model, b) static WS network (figure \ref{fig:ws_spec}),
c) static NW network (figure \ref{fig:nw_spec}), d) dynamic WS network,
e) dynamic NW network. Here the $x$ axis is the time, and the y axis
is the $\log_{2}\left(k_{n}\right)=n-2$. Barriers that we see in
b) are due to nodes with missing connections. It is interesting that
while both (b) and (c) are slower to settle to the steady state than
(a), both (d) and (e) are faster or same. It also seems that (c),
(d) and (e) all have slightly different dynamics from (a) in that
they seem to spend more time with energy localized mainly at large
scales, which appear as blue gaps around $\log_{2}\left(k_{n}\right)\approx10$.
We see these gaps for instance between $t=1000$ and $t=1200$ in
(e) and $t=900$ and $t=1050$ in (d). The equivalent gap we see in
(a) around $t=650$ is much narrower in comparison.}
\end{figure}

The initial phase of the time evolution for different models can be
seen in Figure \ref{fig:evol}. Here, the nodes with less than three
connections act as barriers in the static WS case. This results in
a slower buildup and a very noisy final spectrum as shown in figure
\ref{fig:ws_spec}. In contrast, in the static NW case, non-local
connections weaken the initial broadening of the energy spectrum around
the production region by coupling directly to small scales that are
strongly dissipative. This results in a slower buildup as well. However
since there are no barriers in NW, the final state is roughly the
same with that of GOY. In contrast, since the network evolution time
scale $\Delta t=10^{-2}$ is much faster compared to the time it requires
to reach steady state, evolving network acts as a halo connecting
all the nodes to one another, speeding up the redistribution of energy.
Changing $\Delta t$ has a nontrivial impact on the dynamics of WS,
but not so for NW. Since WS has barriers, how long those stay in one
place affects the dynamics. We don't show a $\Delta t$ scan here
explicitly, but this can be seen from the difference between the static
(i.e. $\Delta t\rightarrow\infty$) vs. dynamic network versions of
the WS shown in figure \ref{fig:evol}.

\begin{figure}
\begin{centering}
\includegraphics[width=0.8\columnwidth]{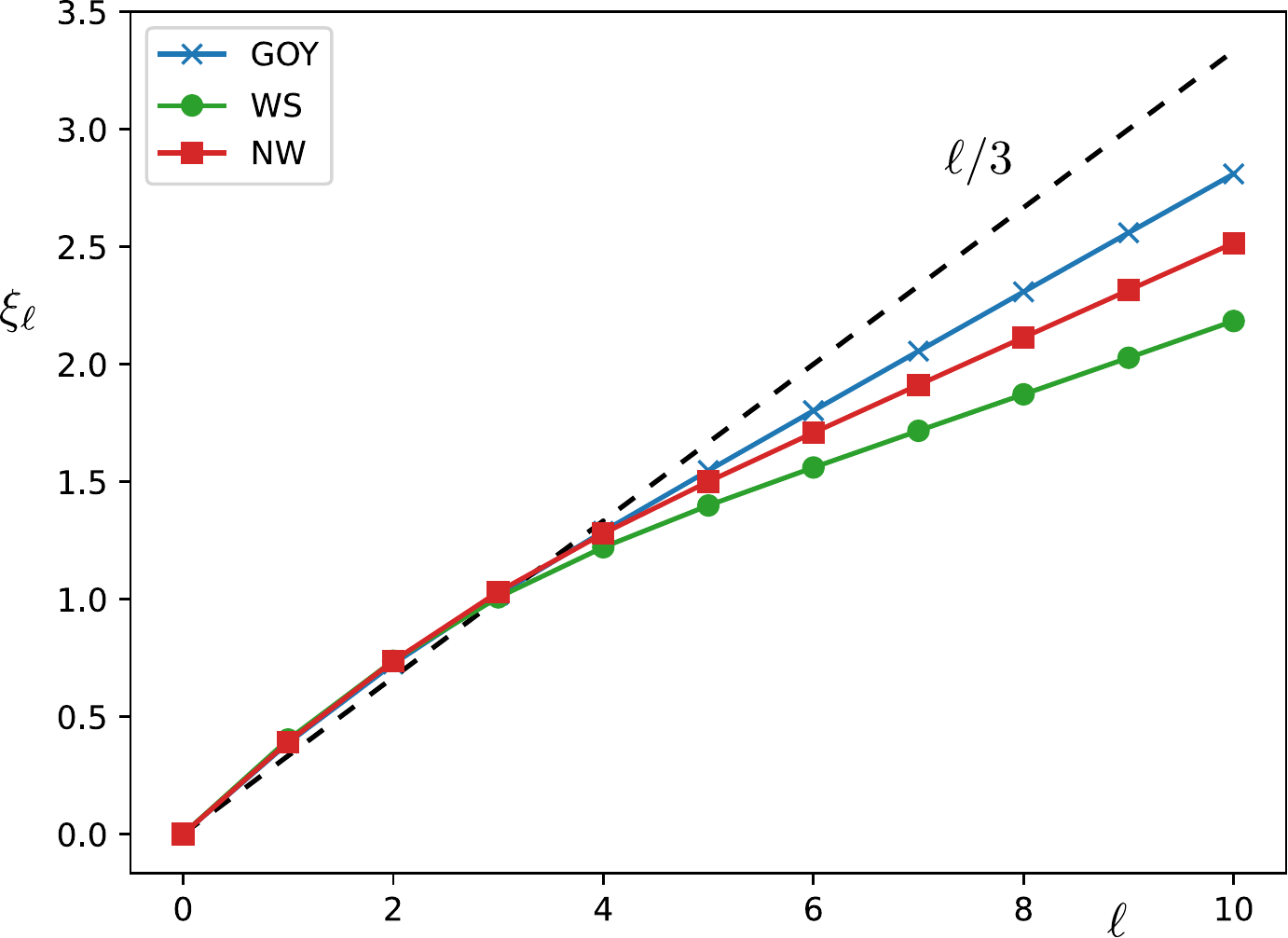}
\par\end{centering}
\caption{\label{fig:inter}Intermittency in dynamical complex network models.
It seems that intermittency increase as the ratio of random non-local
connections to local nearest neighbor connections increase. NW increase
this ratio by adding non-local interactions, whereas WS increase it
further since it also removes local connections as it adds non-local
ones.}
\end{figure}

Another interesting tool in understanding the dynamics of the turbulent
cascade is the structure function, which gives information about the
scale by scale distribution of statistical features of the flow field.
The shell model equivalent of the $\ell$th order structure function
can be written as $S_{\ell}^{n}=\left\langle \left|u_{n}\right|^{\ell}\right\rangle $
where the average is to be computed over time. Assuming that it has
a power law form: $S_{\ell}^{n}\propto k_{n}^{-\xi_{\ell}}$, one
can obtain $\xi_{\ell}$ by considering $y_{\ell}=\log_{10}\left(S_{\ell}^{n}\right)$,
and $x=\log_{10}\left(k_{n}\right)$ and making a linear regression
to obtain $y_{\ell}=a_{\ell}+b_{\ell}x$, so that $\xi_{\ell}\approx-b_{\ell}$.
When we plot this as a function of $\ell$ as in figure \ref{fig:inter},
its deviation from the theoretical estimate, $\xi_{\ell}=\frac{\ell}{3}$
gives us an indication of the intermittency. Somewhat expectedly,
the intermittency increases when the ratio of nonlocal to local connections
increase. 

The GOY model is rather successful in capturing the key features of
intermittency\citep{benzi:93,pisarenko:93}, thought to be due to
instanton dynamics\citep{mailybaev:12}. Therefore including non-local
interactions that rewire randomly increasing its intermittency is
not really very useful. However other models\citep{gurcan:17,gurcan:18,gurcan:19},
more complex than GOY, which can address various aspects of turbulence,
including anisotropy, lack intermittency corrections. It would be
interesting to devise similar modifications for these models.

Note finally that in contrast to the case of stochastically perturbed
shell models\citep{biferale:97} the intermittency increase with random
perturbations of the lattice structure (via the introduction of long
range interactions) without any perturbation on the equations themselves,
the increase of intermittency is rather significant.

\section{Conclusion\label{sec:conc}}

We have introduced complex network models as a generalization of the
GOY model to arbitrary networks, where the shells represent nodes
and the network consist of connections between them. Its structure
can be identified as a bipartite network between wave-number nodes
and triad nodes, where each triad is connected to three different
wave-number nodes. The approach allows to decouple the setting up
or the evolution of the network topology from the evolution of the
node variables $u_{n}\left(t\right)$ on the network.

We have discussed two basic strategies of network wiring based on
replacing existing local interactions by nonlocal ones (WS), or adding
nonlocal interactions (NW) on top of the existing connections. While
static results show an oversized effect of the network topology on
the cascade, dynamically evolving network models show a more statistical
effect.

In fact, when the network is dynamically rewired from an original
regular lattice with a time step $\Delta t$, we find that for $\Delta t\sim\delta t$,
where $\delta t$ is the correlation time of the forcing, we get almost
exactly the same $k$-spectrum but slightly higher intermittency observable
both in terms of temporal dynamics (i.e. appearance of larger gaps
in time evolution), and when it is computed properly using deviation
of the scaling of higher order structure functions from Kolmogorov
theory. We find that in particular for the WS case, how fast the network
evolves plays an important role in both the dynamics and in the final
steady state result. Since WS can have wave-number nodes with a degree
less than $3$, it can produce barriers for the energy cascade, and
how long those barriers remain in one place is detrimental to the
evolution of the spectrum.

Various obvious ideas, such as the use of preferential attachment
strategies that lead to scale-free networks have been left to future
studies. We believe that focusing on the formulation and considering
a few simple strategies allows us to perform a more detailed study
and present a more coherent picture of the connection between turbulence
and networks.

Since it is argued in the introduction that the network topology represents
interaction efficiency of the full turbulent system when represented
as a reduced model, such as GOY, mainly due to phase relations, one
may try to ``extract'' the network structure by computing instantaneous
shell to shell energy transfer in a fully resolved direct numerical
simulation. However, this is an serious undertaking and is therefore
left to future studies.
\begin{acknowledgments}
I would like to thank Prof. W.-C. Müller and Dr. Ö. Gültekin for fruitful
discussions.
\end{acknowledgments}

\end{document}